\documentclass[reprint,superscriptaddress,aps,longbibliography]{revtex4-2}
\usepackage[utf8]{inputenc}
\usepackage[T1]{fontenc}
\usepackage{amsmath}
\usepackage{amssymb}
\usepackage{graphicx}
\usepackage[capitalize, nameinlink]{cleveref}
\usepackage{subfigure}
\graphicspath{{figures/}}
\usepackage{adjustbox}
\usepackage{booktabs}
\usepackage{multirow}

\usepackage{color}
\usepackage{rotating}
\usepackage{ulem}
\usepackage{soul}

\begin{document}

\title{Ferromagnetic resonance in 3D-tilted square artificial spin ices}

\author{Ghanem Alatteili}
\thanks{These authors contributed equally.}
\author{Alison Roxburgh}
\thanks{These authors contributed equally.}
\author{Ezio Iacocca}
\affiliation{Center for Magnetism and Magnetic Nanostructures, University of Colorado Colorado Springs, Colorado Springs, CO 80918 USA}

\date{\today}

\begin{abstract}
Artificial spin ices (ASIs) arranged in square formations have been explored from the perspective of reconfigurable magnonics. A new frontier in ASIs is their three-dimensional (3D) extension. Here, we numerically explore the ferromagnetic resonance of square ASIs as each nanomagnet is rotated out of plane into 3D ASIs, in which the vertex gap can be either kept constant or varying. We study both remanent and vortex configurations using a semi-analytical dynamic approach and micromagnetic simulations. We find that both methods show qualitative agreement of the main spectral features. However, there are important limitations. On one hand, the semi-analytical approach relies on a minimal model of the demag field, preventing exact predictions. On the other hand, micromagnetic simulations suffer from sufficient resolution, making the results grid-dependent and extremely slow. Regardless, both methods display tunability as a function of the tilt angle. These results showcase advantages and limitations of both methods and are promising to further our understanding of 3D ASI dynamics.
\end{abstract}
\maketitle

\section{Introduction}

Artificial spin ices (ASIs) are geometric arrangements of nanomagnets that are coupled by static and dynamic dipole interactions~\cite{Heyderman2013,Skjaervo2020}. These periodic arrays lead to geometric frustration between the nanomagnets~\cite{Nisoli2013}, where we define frustration as the competition between interactions such that they cannot all be satisfied simultaneously~\cite{Wang2006}. Due to the ASIs' geometric frustration, there is a large degeneracy of magnetic states which can be accessed by different magnetic protocols. For this reason, ASIs can be  considered as strong candidates for reconfigurable magnonics~\cite{Gliga2020,Lendinez2019}. 

Different ASI arrangements have been explored, including square~\cite{Wang2006}, Kagome~\cite{Tanaka2006}, toroidal~\cite{Li2022}, trident~\cite{Farhan2017}, Shakti~\cite{Lao2018}, and Santa Fe~\cite{Mondal2024} lattices. An additional classification of ASI configurations is based on their magnetization state. In the case of square ASIs, the ground state is the vortex state and the first ``excited'' state is the remanent state~\cite{Heyderman2013}.

Dynamics in ASIs have been primarily investigated in the square ASI configuration. Numerical simulations, calculations, and experiments demonstrated that local defects had footprints in the FMR spectrum~\cite{Gliga2013} and that the magnonic band structure was intrinsically related to the magnetization states~\cite{Iacocca2016,Jungfleisch2016,Iacocca2017c,Iacocca2020}. More recently, the dynamics have been enriched by lattice modifications~\cite{Arroo2019,Dion2019,Gartside2021,Lendinez2021,Vanstone2022} which has allowed for applications in reservoir computing~\cite{Gartside2022,Saccone2022,Lee2024}; nonlinear dynamics have been achieved in square ASIs~\cite{Lendinez2023}; and ultrastrong dynamic coupling has been measured in a multilayered spin-vortex ASI~\cite{Dion2024}.

Three-dimensional (3D) geometries of ASIs is a recent branch of exploration where significant advances have been made~\cite{Gliga2019,May2019,May2021,Pip2022,Harding2024}. It has been found that 3D ASI geometries mimic the bulk spin-ice dynamics~\cite{Saccone2023b}, offering increased reconfigurability and more controllable magnetostatic interactions. Theoretical studies of 3D geometries have shown the emergence of tensionless Dirac strings and mobile magnetic monopoles that can be tuned via an external magnetic field~\cite{Koraltan2021}. Despite the hurdles of 3D research, including nontrivial characterization and fabrication techniques, magnons have been measured by Brillouin light scattering in a cubic lattice with broken symmetry~\cite{Sahoo2021} and cubic networks~\cite{Guo2023}. Many additional avenues can be investigated based on anisotropy induced by curvilinear magnetism~\cite{Volkov2019} and frustration in 3D ASIs~\cite{Mistonov2013,Rana2023,Koshikawa2023}.

Here, we investigate the FMR resonance of a square ASI configuration where its individual elements are rotated out of the plane of the ASI, forming a tilted square ASI geometry. This tilted configuration opens a new degree of freedom to modulate the coupling between nanomagnets. Both G\ae{}nice~\cite{Alatteili2023}, a semi-analytical eigenvalue solver, and micromagnetic simulations are used. The methods exhibit several qualitative similarities in the predicted spectra but present important differences. For example, G\ae{}nice's treatment of demag fields as a diagonal tensor impacts the accuracy of the coupling while the finite difference approximation of MuMax3~\cite{Vansteenkiste2014} is not optimized for such curved geometries. Nevertheless, our work shows as a first step that FMR modes are tunable with the tilt angle and suggests that investigation of ASI geometries with 3D modifications can exhibit interesting dynamical phenomena in their dispersion relation.

The remainder of the paper is organized as follows: In section II, the geometry is described. The magnetization states computed by micromagnetic simulations are described in section III. Section IV discusses the FMR resonance as a function as the tilt angle both from micromagnetic simulations and our semi-analytical model G\ae{}nice~\cite{Alatteili2023}. Finally, we provide our concluding remarks in section V.

\section{Geometry}

\begin{figure}[t]
\centering
\includegraphics[width=2.5in]{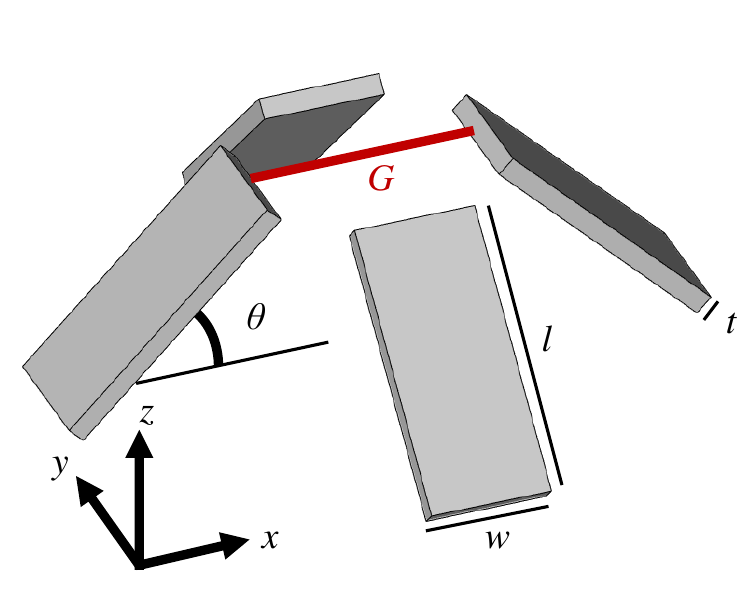}
\caption{Schematic of the tilted ASI unit cell. The nanomagnets are shown here are rectangular prisms of dimensions $l\times w\times t$ for simplicity, but the calculations assume stadium-shaped nanomagnets. The nanomagnets are tilted by an angle $\theta$. The gap distance $G$ is defined as the distance between the tip of two face-to-face nanomagnets. The lattice constant $d$ (not depicted) is the center-to-center distance between nanomagnets, as typically defined for square ASIs.}
\label{fig:schematic}
\end{figure}

In a tilted square ASI, the unit cell is composed of four nanomagnets placed around a vertex. Each nanomagnet is tilted out of the $x-y$ plane by the polar angle $\theta$. We consider here a particular case where the magnets are rotated such that their tips converge at the vertex in order to maximize dipolar coupling. A schematic of the unit cell is shown in Fig.~\ref{fig:schematic}. Each nanomagnet is modeled by the dimensions $l=280$~nm, $w=100$~nm, and $t=20$~nm.

We investigated two possible cases to construct the tilted square ASI: fixing the nanomagnets' center position or fixing the gap between nanomagnets, $G$. These cases are reflected in the translation vectors that construct the lattice, $\mathbf{a}_1=2d$~$\hat{x}$ and $\mathbf{a}_2=d(\hat{x}+\hat{y})$, where $d$ is the center-to-center distance between two adjacent nanomagnet geometric centers. In the case of fixed distance, we choose $d=430$~nm which corresponds to a gap distance in the in-plane configuration ($\theta=0$) of $G=150$~nm. In the case of fixed gap, $d$ varies with respect to the tilt angle $\theta$
\begin{equation}
\label{eq:5}
    d = l\cos{(\theta)} + G.
\end{equation}

Clearly, Eq.~\eqref{eq:5} leads to a vertical arrangement of nanomagnets in the limiting case of $\theta=90$~deg, which would be reminiscent of geometrically arranged nanorods~\cite{Verba2012}. We thus limit our study to $0\leq\theta\leq80$~deg.

Throughout this paper, we consider permalloy parameters $M_{s} = 790$~kA/m, $A = 10$~pJ/m, $\alpha=0.01$, and negligible magnetocrystalline anisotropy.

\section{Static magnetic configurations}

We modeled the tilted square ASI using MuMax3~\cite{Vansteenkiste2014}. Individual magnetic configurations were obtained by simulating each ASI at a specific angle $\theta$, in increments of 2.5 degrees from 0 degrees to 80 degrees. Each ASI was set with periodic boundary conditions in the $x$ and $y$ directions. For the case of fixed distance, the simulation domain was a cube of sides $2d$ so that it supported cubic cells and was sufficient to capture the unit cell and their immediate neighbors. Therefore, the simulation domain had dimensions $860$~nm~$\times~860$~nm~$\times~860$~nm discretized in $256$~$\times~256$~$\times~256$ cells so that the cubic cell had sides of length $3.36$~nm. For the case where the gap is fixed to $G=150$~nm, the simulation domain was allowed to vary according to Eq.~\eqref{eq:5}, while keeping the number of cells constant. This means that the minimum cell size was achieved when $\theta=80$~deg, in which case the cube had size $1.55$~nm.

\begin{figure*}[t]
\begin{center}
\includegraphics[scale=0.75]{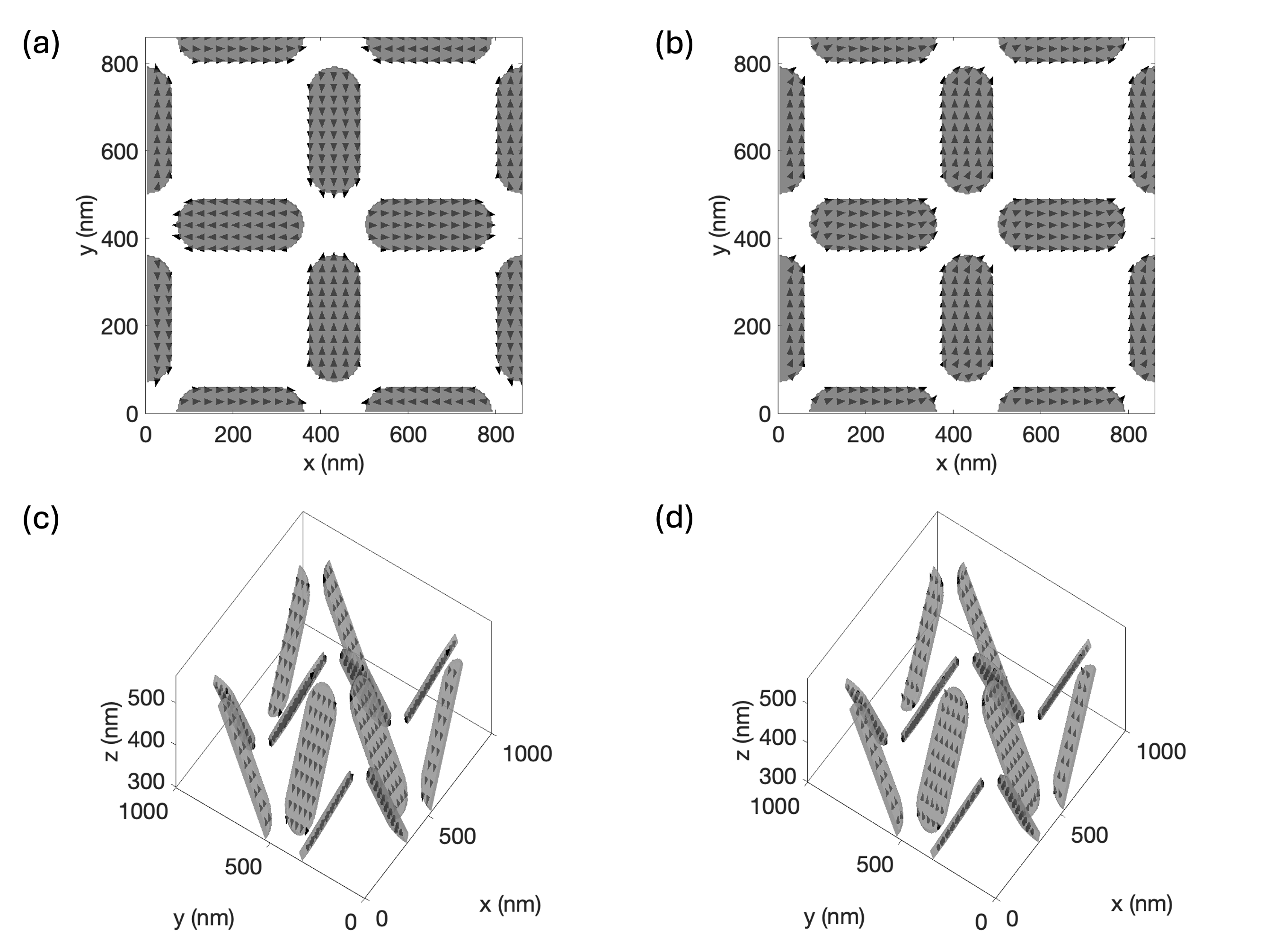}
\caption{Magnetization at $\theta=0$ for the (a) vortex state and (b) remanent state, and for $\theta=45$ in the (c) vortex state and (d) remanent state. The arrows represent the magnetization orientation.}
\label{RemFGVortexFG_Vis_N256}
\end{center}
\end{figure*}

For each case, we stabilize both the magnetic ground state, or vortex state, and the remanent state. The stabilized magnetic states for $\theta=0$ are shown in Fig.~\ref{RemFGVortexFG_Vis_N256}, displaying the (a) vortex and (b) remanent state. This simulation is our benchmark and illustrates the well-known result that bending of the magnetization at the edges is observed in the remanent case~\cite{Gliga2015}, contributing to the higher energy of this configuration compared to the vortex state. We repeat this procedure for both fixed distance and fixed gap cases. As an illustrative example, we show in Fig.~\ref{RemFGVortexFG_Vis_N256} the stabilized states for $\theta=45$~deg in the fixed gap case, displaying the (c) vortex and (d) remanent state. Similar to the $\theta=0$ case, there is visible edge bending only in the remanent configuration. From these simulations, we can estimate the edge bending, $\Phi$, as a function of $\theta$. The obtained angles are shown in the appendix~\ref{appang}.

\section{Ferromagnetic resonance}
 \begin{figure*}[t]
 \begin{center}
 \includegraphics[scale=0.75]{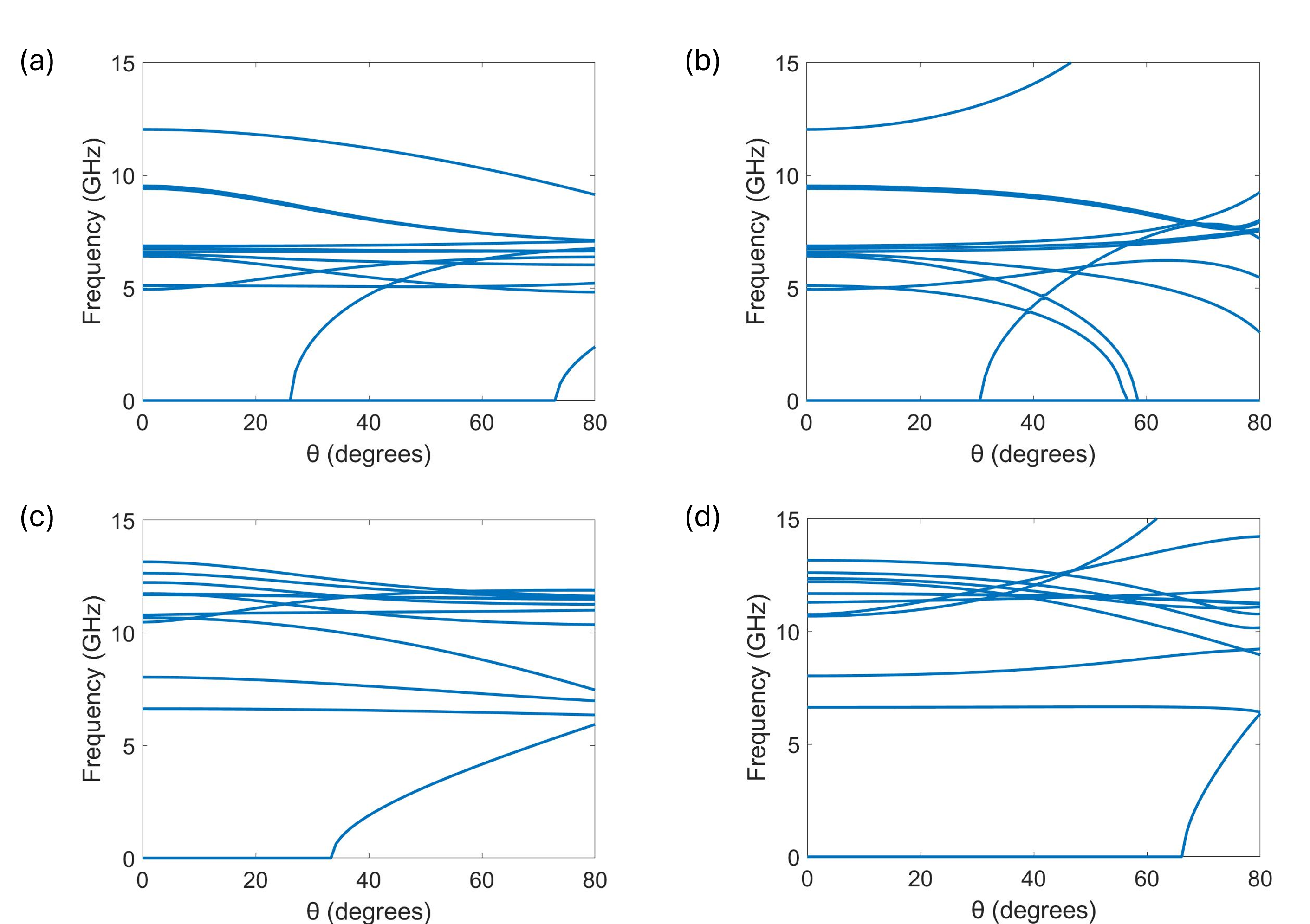}
 \caption{Ferromagnetic resonance computed with G\ae{}nice as a function of the tilt angle $\theta$ ranging from $0$ to $80$~degrees for the following cases: (a) with a fixed distance and (b) with fixed gap in the vortex configuration, (c) with fixed distance and (d) with the fixed gap in the remanent configuration.}
 \label{Dispersions}
 \end{center}
 \end{figure*}
 
The FMR of the tilted square ASIs is computed by means of G\ae{}nice, a semi-analytical approach that solves the linear eigenvalue problem of arbitrary ASI configurations~\cite{Alatteili2023}. The formalism in G\ae{}nice is based on the Holstein–Primakoff transformation for small amplitude excitations~\cite{Slavin2009} to construct a Hamiltonian consisting of a uniform external magnetic field, shape anisotropy using a diagonal tensor approximation~\cite{Martinez2023}, exchange interaction, and dipole-dipole interaction. G\ae{}nice uses a tight-binding model for dipole-dipole interactions in order to simplify the Hamiltonian. The nanomagnets are split into three macrospins to capture the exchange energy contribution to the dispersion of predominantly dipolar waves. This approximation results in a significant computational speed at the expense of some numerical accuracy. So far, G\ae{}nice has demonstrated good agreement with experiments in trilayered square ASIs~\cite{Dion2024}.


The computed FMR as a function of the tilt angle $\theta$ are shown in Fig.~\ref{Dispersions}. The FMR for the vortex state is shown in (a) and (b) for the fixed distance and fixed gaps cases, respectively. For these calculations, we assumed demag factors from an ellipsoid~\cite{Osborn1945} for simplicity. In both cases, we resolve modes that are split at $\theta=0$, with the highest frequency being a bulk mode and the lowest frequency being an edge mode. The most notable difference between both cases is that the highest frequency mode redshifts for the fixed distance case, panel (a), and blueshifts for the fixed gap case, panel (b). We attribute this behavior to the static dipole field that becomes weaker in the fixed distance case, thus reducing the net Zeeman energy on each nanomagnet. Additionally, the modes at approximately $5$~GHz and $7$~GHz split when the planar symmetry is broken, $\theta\neq0$. While both red- and blueshifts occur as the bands split, there is a much stronger dependence in the fixed gap case, with two modes softening close to $\theta=60$~deg. This softening is merely a consequence of the linear approximation and there is no indication of destabilization of the magnetization state from micromagnetic simulations. Finally, it is worth noting two similarities between panels (a) and (b). The mode at $\approx9$~GHz redshifts in both cases, suggesting that this mode is primarily due to exchange. In fact, experimental results often show a secondary peak which has has been reproduced with our semi-analytical formalism~\cite{Jungfleisch2016,Dion2024} which is not harmonically related to the fundamental peak. These results indicate that such a peak would exhibit a different qualitative trend with $\theta$. In addition, a new mode appears at $30$~deg which is an indication of further symmetry breaking in the system leading to real eigenvalues.

In the remanent state, the fixed distance and fixed gap cases are shown in Fig.~\ref{Dispersions}(c) and (d), respectively. The magnetization's edge bending found with micromagnetic simulations returns a nonlinear behavior. We perform polynomial fits to the angles estimated from micromagnetic simulations as shown in the appendix~\ref{appang}. These are approximations that certainly compromise the accuracy of the computed eigenvalues, as discussed later. The computation of FMR modes with the $\theta$-dependent edge bending reveals a notable redshift in both states, particularly at higher frequency modes shown in Figures ~\ref{Dispersions}(c) and ~\ref{Dispersions}(d). While more modes are available here, we can discern a few modes blueshift in the fixed gap case, similar to the vortex scenario. Additionally, in the fixed distance case, four modes converge towards $\approx7$~GHz as a function of $\theta$. This is again due to the reduced static dipole field as the nanomagnets are further apart.


\begin{figure*}[t]
\begin{center}
\includegraphics[scale=0.75]
{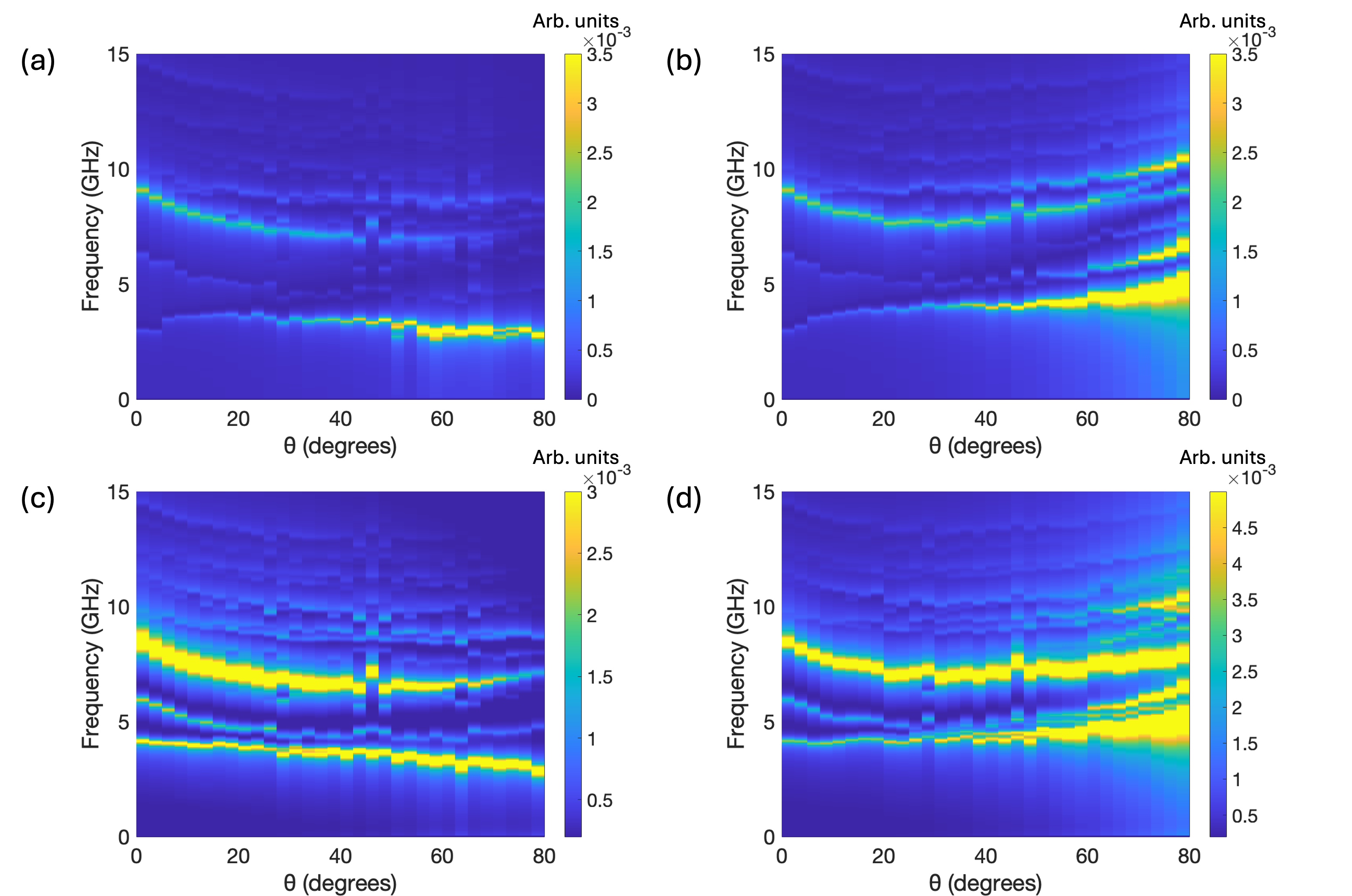}
\caption{Micromagnetic spectra (a) with a fixed distance and (b) with a fixed gap between the ASIs in the vortex configuration, then (c) with a fixed distance and (d) with a fixed gap in the remanent configuration. The colorbar displays the spectral intensity in arbitrary units.}
\label{Combined_Vortex_Spectra_MuMax}
\end{center}
\end{figure*}


These calculations provide insights into the available modes in the system within a minimal model. However, the intensity of each mode is not retrieved from the calculation in the present implementation of G\ae{}nice. Typically, only two modes are dominant in experiment, bulk and edge mode. To determine whether the minimal model is representative of the expected FMR, we compare our results with those obtained by micromagnetic simulations.

We retrieve FMR from MuMax3 according to the following procedure. First, we avoid waves originating from the discretization by performing a second relaxation of the stabilized magnetic states with a realistic damping of $\alpha=0.01$ for $15$~ns. After this time, we apply a uniform external field pulse of $10$~mT in the $z$ direction for a time of $10$~ps, and then allow the simulation to relax for $20$~ns, sampling at $10$~ps. These parameters return a frequency resolution of $50$~MHz and access a maximum frequency of $50$~GHz.

The micromagnetic spectra are obtained by averaging the individual spectra for each magnetization component. This was done to take into account the tilt angle relative to the field pulse. These results are displayed in Fig.~\ref{Combined_Vortex_Spectra_MuMax} for the same cases explored with G\ae{}nice, namely, vortex state in panels (a) and (b), and remanent state in panels (c) and (d), each for the cases of fixed distance and fixed gap, respectively. A common feature of all spectra is the redshift of the high-frequency mode at angles below $\theta=30$~deg. However, the low-frequency mode is observed to blueshift for the vortex state and redshift for the remanent states. These features were not reproduced by G\ae{}nice as discussed later.

At larger tilt angles, there is a higher number of observable modes, all of which appear to blueshift. The blueshift is visibly larger for the fixed gap cases, panels (c) and (d). In particular, the lowest-frequency mode experiences the largest changes for the fixed distance in the remanent state, panel (c), where it redshifts from 4~GHz to $3$~GHz; and the fixed gap case in the vortex state, panel (b), where it blueshifts from 4~GHz to 5~GHz. Another notable difference between the fixed distance and fixed gap cases is that the higher frequency bands achieve larger intensity when $\theta>70$~deg. This is likely due to coupling between the excitation pulse and the edge modes.

\begin{figure}[t]
\begin{center}
\includegraphics[scale=0.75]
{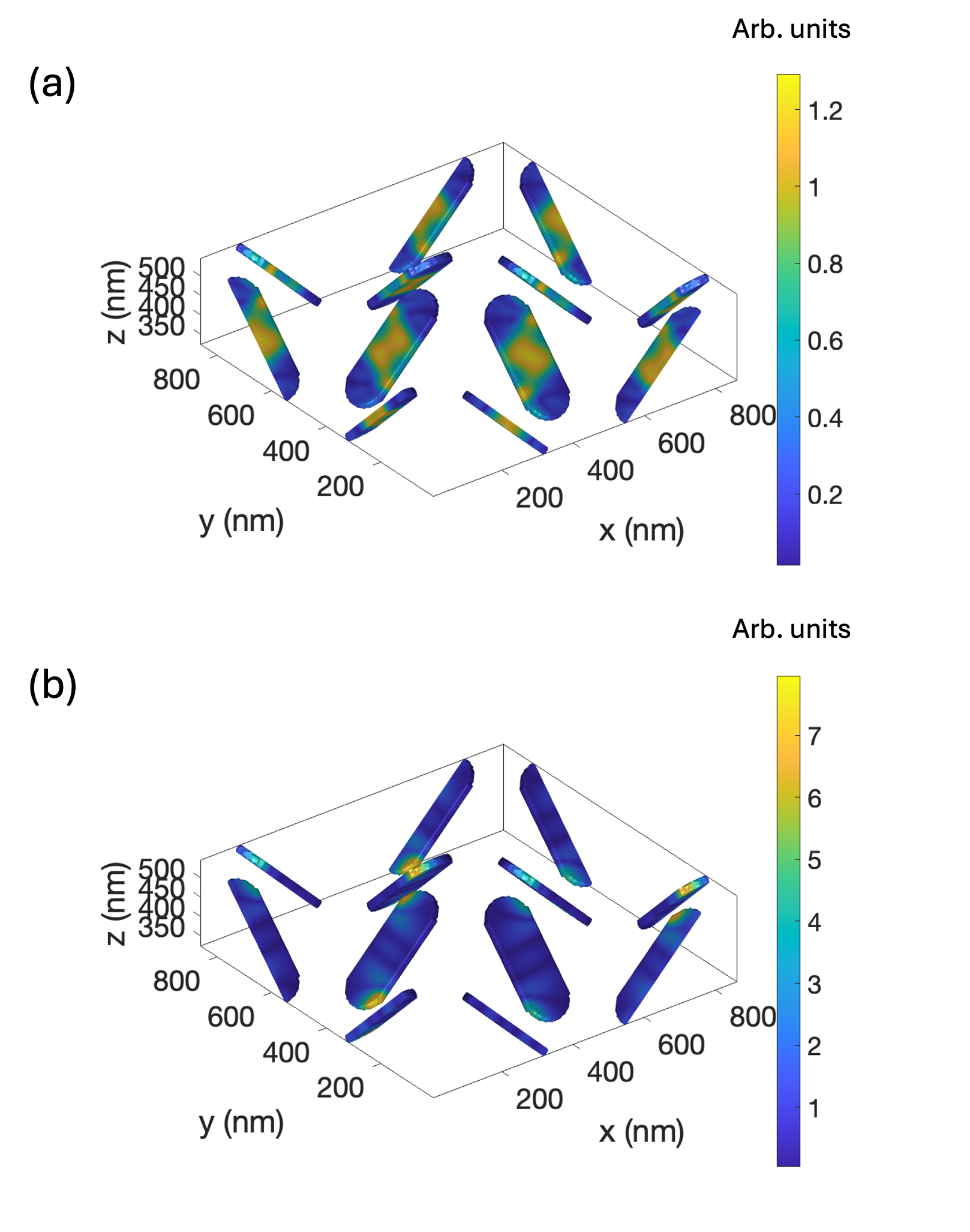}
\caption{Modes of the ASIs tilted to $\theta=45$~deg in the remanent configuration with a fixed distance at a frequency of (a) 6.18~GHz displaying the bulk mode and (b) 3.78~GHz displaying the edge modes. The colorbar displays spectral intensity in arbitrary units.}
\label{ModesFig}
\end{center}
\end{figure}

Micromagnetic simulations also allow us to determine the spatial extent of the modes, confirming that the highest-frequency mode is a bulk mode and the lower-frequency mode is an edge mode. An example is shown in Fig.~\ref{ModesFig} for the remanent state in the fixed distance case at $\theta=45$~deg. The bulk mode at $6.18$~GHz is shown in panel (a) where the spectral intensity is more prominent within the nanomagnets. The edge mode, displayed in panel (b), has a frequency of 3.78~GHz and exhibits a higher intensity at the edges of the nanomagnets.

We observed indications of qualitative agreement between the G\ae{}nice model in Fig.~\ref{Dispersions} and the micromagnetic simulations in Fig.~\ref{Combined_Vortex_Spectra_MuMax}. Both models show a respective redshift and blueshift of the high-frequency mode which we attribute to the reduced and increased stray field in the fixed distance and fixed gap cases, respectively. G\ae{}nice also predicts frequency split due to the symmetry broken by the 3D tilt and thus the increase of spectral content as $\theta$ increases, which is also observed in micromagnetic simulations for most cases, the exception being the fixed distance vortex state. In this case, the modes' split is not significant in G\ae{}nice and not observed in micromagnetic simulations. 

However, there are important discrepancies, particularly a significant drop in frequencies at low tilt angles, in the $0<\theta<20$~deg range in the micromagnetic simulations, shown in Fig.~\ref{Combined_Vortex_Spectra_MuMax}, that is not observed in the G\ae{}nice calculations. We attribute this discrepancy to the particularities of the demag field calculation which is resolved numerically in micromagnetic simulations but is approximated by a diagonal tensor in G\ae{}nice~\cite{Alatteili2023}. In other words, G\ae{}nice does not resolve the details of the demag field due to the nanomagnet's shape. However, we note that the diagonal approximation can be fitted to numerical FMR calculations~\cite{Martinez2023} to improve accuracy. We also note the emergence of low-frequency modes in G\ae{}nice  at tilt angles above 40~$\deg$, as shown in Fig.~\ref{Dispersions}, that are not observed in the micromagnetic simulations. These soft modes are likely another consequence of the diagonal approximation of the demag field.



\section{Conclusions}

We computationally explored the FMR of tilted square ASIs as a function of tilt angle, spacing, and magnetization configurations. We compared results from two different methods, the first being a semi-analytical approach, G\ae{}nice, and the second by micromagnetic simulations. Both models show qualitative agreement, recovering the redshift and blueshift for the bulk modes in the fixed distance and fixed gap cases, respectively. 

The quantitative differences between the two methods are due to several limitations. On one hand, G\ae{}nice relies on approximations for computational efficiency, such as the discretization of nanomagnets into three-macrospins and a diagonal approximation for the demag field. In addition, the magnetization edge bending in the remanent state is not actively computed in G\ae{}nice because it is an eigenvalue solver. An improvement would allow G\ae{}nice to perform energy minimization to resolve edge bending due to magnetostatic interactions within the minimal model. On the other hand, micromagnetic simulations using Mumax3~\cite{Vansteenkiste2014} are based on a finite difference approach so that the discretization is crucial for the correct calculation of the demag factor and resolution of ferromagnetic resonance. In a 3D geometry, this approach becomes computationally expensive and the accuracy of the calculations is not guaranteed, requiring instead finite elements implementations~\cite{Cheeninkundil2023,Golebiewski2024}. Additionally, micromagnetic simulations require the definition of a multitude of static states at individual angles for each magnetization configuration and spacing variation, making investigation of 3D geometries more time consuming than G\ae{}nice's analysis of dynamic cases. 

This paper reports FMR for a 3D ASI and demonstrates the tunability of these frequency modes with varying tilt angles, highlighting the geometrical optimization offered by varying ASI configurations. While these structures are difficult to realize experimentally, our results comprise a first exploration of 3D ASI relying predominantly on a semi-analytical model that can allow optimization of geometrical parameters prior to a more detailed exploration with micromagnetic simulations and even experiments.

\section*{Acknowledgments}

This material is based upon work supported by the National Science Foundation under Grant No. 2205796.

\appendix
\section{Magnetization edge bending}
\label{appang}

\begin{figure*}[t]
 \begin{center}
 \includegraphics[scale=0.45]{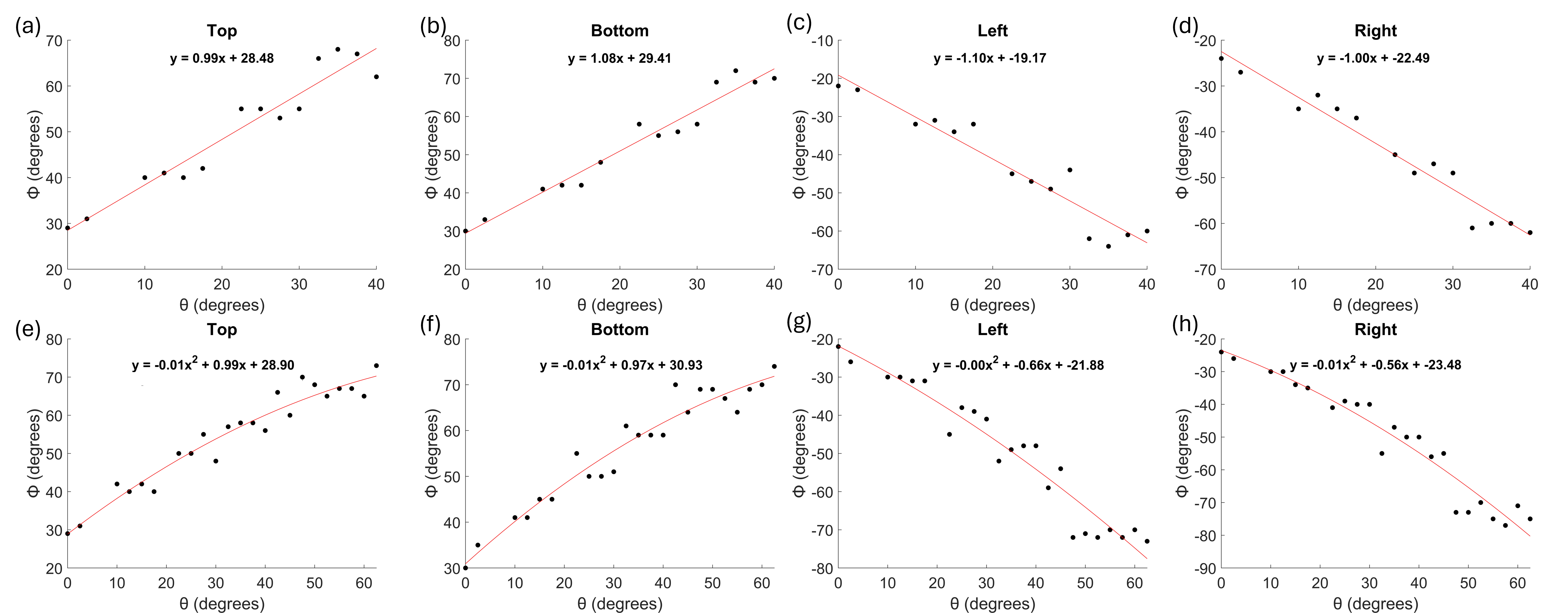}
 \caption{ The relationship between the remanent edge magnetization $\phi$ and the tilt angle $\theta$ for the (a) top, (b) bottom, (c) left and (d) bottom bending measurement in the fixed distance case and  (e) top, (f) bottom, (g) left and (h) bottom bending measurement in the fixed gap case. The plot displays the measurements, represented by filled black circles. A linear fit is superimposed in red for the fixed distance case, while a second-order polynomial fit is shown for the fixed gap case. The fit equation for each case is displayed in bold in each panel. }
 \label{fig:Fits}
 \end{center}
 \end{figure*}

The magnetization edge bending is estimated from micromagnetic simulations as an average angle $\Phi$ at the edges. Because the simulation is 3D and the extent of the ``edge'' depends on the mode volume, these estimates are rather crude and calculated on a case-by-case basis by inspection. The main goal here is to provide G\ae{}nice with an approximate edge bending function of the tilt angle $\theta$.

The results from micromagnetic simulations are shown in Fig.~\ref{fig:Fits} by black circles. The top row (a-d) represents the fixed distance case and the bottom row (e-h) the fixed gap case. We distinguish the edge utilized by their location relative to the geometric center of the micromagnetic cell. Clearly, the evolution of these angles is nonlinear, despite the crude estimation. We perform polynomial fits to each case and find acceptable descriptions with linear fits for the fixed distance case and second-order polynomials for the fixed gap case.

\end{document}